\documentclass[%
 reprint,
superscriptaddress,
 amsmath,amssymb,
 aps,
]{revtex4-2}

\usepackage{graphicx}
\usepackage{dcolumn}
\usepackage{bm}

\usepackage{enumitem}
\usepackage[caption=false]{subfig}
\usepackage{dsfont}
\usepackage{accents}
\usepackage{hyperref}
\usepackage{cleveref}

\usepackage{tikz}
{
        \end{scope}%
  \pgfresetboundingbox
  \path[use as bounding box] (image.south west) rectangle (image.north east);
  \end{tikzpicture}%
}

\Crefname{figure}{Fig.}{figures}
\Crefname{equation}{Eq.}{equations}
\creflabelformat{equation}{#2\textup{#1}#3}

\begin{document}

\preprint{APS/123-QED}

\title{Experimental observations of fractal landscape dynamics in a dense emulsion}

\author{Clary Rodr{\'i}guez-Cruz}
\affiliation{Department of Chemical and Biomolecular Engineering, University of Pennsylvania, \\ Philadelphia, Pennsylvania}

\author{Mehdi Molaei}
\affiliation{Department of Chemical and Biomolecular Engineering, University of Pennsylvania, \\ Philadelphia, Pennsylvania}

\author{Amruthesh Thirumalaiswamy}
\affiliation{Department of Chemical and Biomolecular Engineering, University of Pennsylvania, \\ Philadelphia, Pennsylvania}

\author{Klebert Feitosa}
\affiliation{Department of Physics and Astronomy, James Madison University; Harrisonburg, Virginia}

\author{Vinothan N. Manoharan}
\affiliation{Harvard John A. Paulson School of Engineering and Applied Sciences, Harvard University, Cambridge, Massachusetts}
\affiliation{Department of Physics, Harvard University, Cambridge, Massachusetts}

\author{Shankar Sivarajan}
\affiliation{Department of Physics and Astronomy, Johns Hopkins University, Baltimore, Maryland}

\author{Daniel H. Reich}
\affiliation{Department of Physics and Astronomy, Johns Hopkins University, Baltimore, Maryland}

\author{Robert A. Riggleman}
\affiliation{Department of Chemical and Biomolecular Engineering, University of Pennsylvania, \\ Philadelphia, Pennsylvania}

\author{John C. Crocker}%
 \email{jcrocker@seas.upenn.edu}
\affiliation{Department of Chemical and Biomolecular Engineering, University of Pennsylvania, \\ Philadelphia, Pennsylvania}

\date{\today}

\begin{abstract}
Many soft and biological materials display so-called ‘soft glassy’ dynamics; their constituents undergo anomalous random motions and complex cooperative rearrangements. A recent simulation model of one soft glassy material, a coarsening foam, suggested that the random motions of its bubbles are due to the system configuration moving over a fractal energy landscape in high-dimensional space. Here we show that the salient geometrical features of such high-dimensional fractal landscapes can be explored and reliably quantified, using empirical trajectory data from many degrees of freedom, in a model-free manner. For a mayonnaise-like dense emulsion, analysis of the observed trajectories of oil droplets quantitatively reproduces the high-dimensional fractal geometry of the configuration path and its associated energy minima generated using a computational model.  That geometry in turn drives the droplets' complex random motion observed in real space. Our results indicate that experimental studies can elucidate whether the similar dynamics in different soft and biological materials may also be due to fractal landscape dynamics.
\end{abstract}

\maketitle

\section{\label{sec:Introduction}Introduction}
Despite the deterministic nature of classical physics, the world around us appears filled with random motion. The random Brownian motion of microscopic particles is due to `noise' -- the collisions of molecules in incessant thermal motion \cite{renn2005einstein}. The random motion of weather systems has a different origin, deterministic chaos, due to the dynamical evolution of their unstable equations of motion \cite{schuster2006deterministic}. The random motion of foraging animals forms a third, distinct type of random motion, a L\'evy walk, containing occasional large displacements \cite{zaburdaev2015levy}. The focus of this work is a class of systems, including foams, emulsions, pastes, and cytoskeletal structures that display random motions having similar mathematical structure, called soft glassy dynamics \cite{gopal1995nonlinear, hebraud1998mode, sollich1997rheology, fabry2001scaling, hoffman2009cell}, which appear to be due to neither thermal fluctuations nor deterministic chaos. These systems have in common strongly interacting and slowly changing constituents forming a disordered solid \cite{corwin2005structural, van2009jamming, morse2017echoes, vasisht2018rate, giavazzi2020multiple, song2022microscopic}, which display super-diffusive motion, non-Gaussian random displacements, and intermittent cooperative motion or `avalanches'. 

A $2016$ simulation study \cite{hwang2016understanding} of a foam was able to reproduce the major features of soft glassy dynamics with a remarkably simple model, and provided insights into its physical and mathematical origins. The model treated the bubbles as frictionless, compressible spheres with no inertia and no thermal noise, whose radii slowly changed to mimic gas diffusion between real bubbles. The bubbles' positions evolved simply according to the minimization of the total system energy.  This corresponds to the system's configuration moving downhill on a potential energy landscape \cite{PELWales2007, lois2010protein, charbonneau2014fractal} that spans a high-dimensional space of all droplet coordinates. The system hopped between minima in this landscape because stable energy minima were occasionally destabilized by the slowly changing bubble radii. 

Analysis of the foam simulation results revealed an unusual and complex geometry for the foam energy landscape and the arrangement of its minima. Interestingly, the random \textit{dynamics} of the foams' bubbles were closely related to features of the fractal \textit{geometry} of the energy landscape traversed by the configuration. The usefulness of such a \textit{fractal landscape dynamics} approach to understanding the physics of foams and emulsions, however, remains untested in experiment. While multi-particle tracking experiments should in principle allow the system's path through a high-dimensional configuration space to be followed, the effects of finite spatial and temporal resolution limit such examination. Moreover, it is not clear how to best characterize the energy landscape geometry using solely positional data and in the absence of energy or stress data, which are available in the simulations.

Here we show how relevant features of the configuration space trajectory, the energy landscape and its minima can be determined in a model-free way using multi-particle tracking data for an index- and density-matched dense emulsion. We find that the measured geometry closely matches the predictions of previous and new simulations, despite finite sampling speed, when straightforward corrections for measurement error are applied. Overall, we find that the droplets' super-diffusive exponent and power-law rheology exponents are related to the configuration space path's fractal dimension, that the non-Gaussian particle displacements are related to the non-random displacement directions taken by the configuration path, and that fractal clustering of energy minima along the path gives rise to power-law distributed avalanche sizes. Our experimental approach may prove useful for studying different systems that have similar dynamics but lack a corresponding simulation model, and whose origins currently defy understanding.

\section{\label{sec:Approach}Empirical Approach for high-dimensional landscapes}
While studying the shape of a fractal curve in a configuration space having hundreds of dimensions may seem daunting, we use three readily understandable geometrical analyses, sketched in Fig. 1. Each of these geometrical features relates to one or more phenomena of soft glassy materials. First, we assess the tortuosity of the configuration space path on different length-scales. As shown in Fig. 1a, we consider random pairs of points on the high-dimensional configuration space path, $\vec{R(}t)$, and compute both the high-dimensional Euclidean distance between them, $\Delta R(t,\tau) = ||\vec{R}(t+\tau)-\vec{R}(t)||$, and the contour distance (or path length) between them, $\Delta s$. Comparing these two distances (averaged over many pairs of points) reports how tortuous the curve is, often quantified with a fractal dimension, $D_f$. Second, we consider the directions taken by the path as it meanders through space. As in Fig. 1b, by studying the angular distribution of the path directions we will determine if the directions are random (isotropic in configuration space), or restricted to a smaller range of directions. Third, we examine the clustering of the energy minima (where the system is in mechanical equilibrium) that the path passes through. Specifically, as in Fig. 1c, we will measure the distribution of Euclidean and contour distances between consecutive minima, $P(\Delta R)$ and $P(\Delta s)$ respectively.  If the minima are clustered into a fractal pattern themselves, these distributions will show a power-law form. Together, these measures provide useful measures of the high-dimensional fractal geometry of the configuration space path and its minima, which in turn give rise to many of the unique phenomena observed in soft-glassy materials.

\begin{figure}[t]
  \centering
  \includegraphics[scale=1]{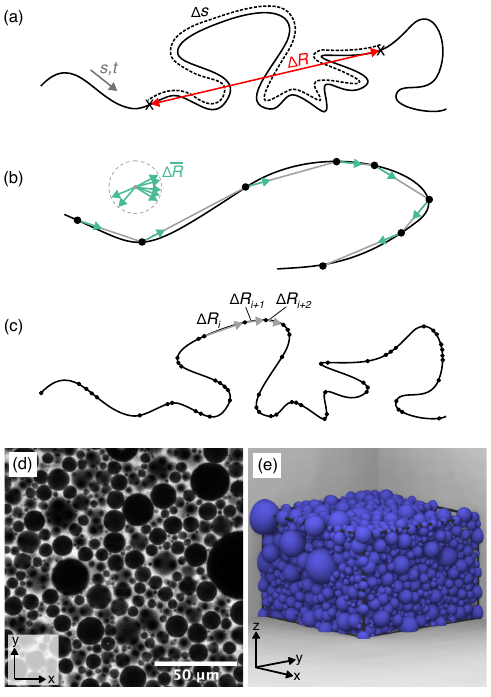}
  \caption{Paths taken by an SGM system can be analyzed experimentally in high-dimensional space. (a) Random pairs of points on the high-dimensional path are chosen to compute their Euclidean distance, $\Delta R$, and their contour distance, $\Delta s$. (b) Displacements between different points are converted to unit length vectors to study their angular distribution. (c) Euclidean distances between adjacent minima (dots) are measured to quantify their spatial distribution. (d) A confocal micrograph shows a section through a dense emulsion, field is 150 $\mu$m on a side. (e) A computer generated reconstruction of the same dense emulsion. Viewing volume is 145x145x100 $\mu$m$^3$}
  \label{fig:cartoon}
\end{figure}

Following the configuration of a soft glassy material through its high-dimensional configuration space requires dynamical tracking of the droplets in three dimensions. We formulated a transparent oil-in-water dense emulsion by matching the oil droplets' index of refraction to that of the aqueous continuous phase. To minimize gravity effects, we also matched the mass density of the two phases to roughly 1 part in 1000. This was achieved by using four liquids, two non-polar ones (1-bromohexane and octane) to make up the droplets (dispersed phase), and two polar ones (formamide and water) to form the continuous phase. The droplets were stabilized with a polymeric surfactant to prevent coalescence. The emulsion was prepared using a commercial homogenizer, with the volume fraction of droplets ($\phi \approx 0.80$) slightly above the jamming threshold, giving it a mayonnaise-like consistency.  

Four dimensional ($xyzt$) imaging of the emulsion in a sealed chamber was performed using a high-speed laser-scanning confocal microscope, imaging fluorescein dye dissolved in the continuous phase. A typical 2-d image is shown in Fig. 1d. The time-dependent droplet positions and radii were determined using multi-particle tracking via custom-written software \cite{clara2015affine, penfold2006quantitative}. A reconstruction is shown in Fig. 1e. Such measurements involve multiple trade-offs; higher magnification and slower scanning result in better spatial location of droplet centers \cite{savin2005static}, but poorer statistical power due to tracking fewer droplets less frequently and associated tracking limitations \cite{martin2002apparent}. Our experiments track 775 droplets to a location accuracy of $\sigma = 0.03\mu$m with two minutes between 3-d scans, (see Appendix C). Our corrections for this finite resolution are discussed in Sections 2 and 4.

As they age, foams and dense emulsions evolve to a steady state termed dynamical scaling \cite{stevenson2010inter} where the shape of the droplet size distribution becomes independent of time, while the mean droplet size increases as a function of sample age. Our samples were allowed to age for 7 hours prior to data acquisition, allowing the system to reach dynamical scaling \cite{feitosa2006bubble} (Appendix B) and slow down to the point that the droplets’ motion was easily followed. Data was collected over a 150 minute span, during which the system could be approximated to have stationary dynamics. 
The experimental results were compared to a simulation using a previously published approach \cite{hwang2016understanding} based upon frictionless, compressible spheres \cite{durian1995foam, tewari1999statistics, ono2003velocity} whose radii slowly evolve due to quasi-static ripening. See Appendix A for further details of the experimental setup and simulation.

\begin{figure}
  \centering
  \includegraphics[scale=1]{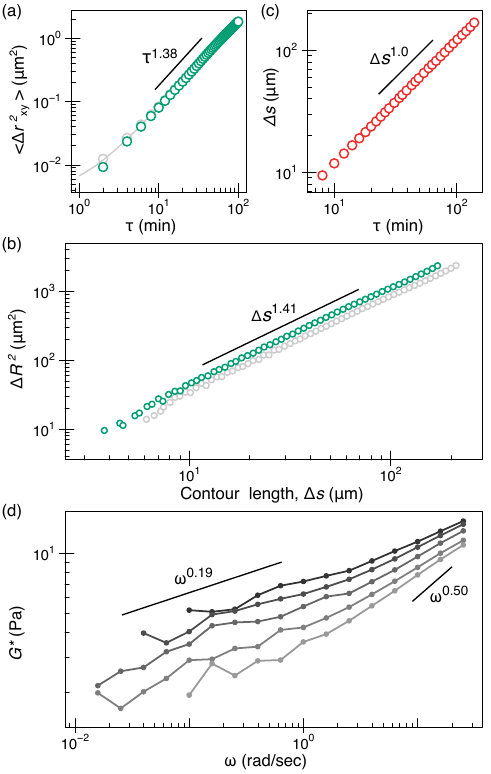}
  \caption{Analysis of configuration space paths and bulk rheology. (a) Mean-squared displacement (green) of individual droplets. Grey symbols are the data before subtracting error and the grey line is its fit to a power law plus a constant (see Appendix C). (b) Squared Euclidean  and contour distances between pairs of configurations show a fractal scaling after smoothing (green) with a slope of 1.41 from a power law fit (for $\Delta s > 20$ $\mu$m). Values without correction for measurement error (grey) show slight deviations from a power law for smaller values. (c) The contour and temporal difference between pairs of configurations shows linear scaling, despite the system’s intermittent dynamics. (d) Measurements of $G^*(\omega)$ show power-law viscoelasticity at low frequencies, whose exponent agrees mathematically with the observed fractal dimension. Samples were loaded 7 hours after formation to correspond to the imaging study, then measured, top to bottom, 3, 10, 40, 75, 127 minutes later.}
  \label{fig:fractal}
\end{figure}

\section{\label{sec:fractal}Super-diffusion and viscoelasticity due to fractal paths}

The random motion of the droplets in our dense emulsion is intermittent. The droplets are nearly motionless except for abrupt motions, or avalanches, where many droplets move by a fraction of their radius, see Movie A1. To quantify such random motion, we will first compute the droplets' mean-squared displacement (MSD), $\langle \Delta r^2(\tau) \rangle = \langle \Delta x^2(\tau) \rangle + \langle \Delta y^2(\tau) \rangle$, where $x$ and $y$ are droplet positions in the horizontal plane, $\tau$ is the lag (or waiting) time and $\langle . \rangle$ denotes an average over multiple droplets and time. Except where noted otherwise, we consider only the $x$ and $y$ coordinates because of their lower measurement error without loss of generality, assuming our system is isotropic. The observed MSD has a super-diffusive form, $\langle \Delta r^2(\tau) \rangle \sim \tau^a$, with $a = 1.38 \pm 0.02$, Fig. 2a. Measurement error affects the MSD by adding a constant noise term of $2\sigma^2$ at short times\cite{martin2002apparent}, which was subtracted from the data and shows $\sigma = 0.03\mu m$ (Appendix C). The physical origin of such power-law super-diffusive motion is not obvious.  Other properties of the bubble motion (discussed in a later section) are inconsistent with existing models for super-diffusion, such as L\'evy walks or chaotic advection \cite{zaburdaev2015levy,solomon1993observation, weeks1998anomalous}, or fractional Brownian motion \cite{mandelbrot1968fractional}. We do find that smaller droplets move faster than larger droplets (Appendix B), consistent with the material acting as a mechanical continuum driven by active fluctuating stresses \cite{lau2003microrheology}.

In the fractal landscape dynamics picture, the steepest descent paths on the energy landscape have a fractal geometry, and this causes the super-diffusion seen in real space\cite{hwang2016understanding}. To test this idea with empirical data, we consider a $1550$-dimensional path $\vec{R(}t)$ constructed from the experimental $x(t)$ and $y(t)$ coordinates of all the droplets. Then, as sketched in Fig. 1a, we consider ‘fragments’ of the path spanning all pairs of observed configurations, and compute the ‘size’ $\Delta R$ and ‘mass’ $\Delta s$ of each path fragment. Similar to the MSD, these high-dimensional quantities are affected by measurement error, but correcting such measurements has not been previously reported. We note that large high-dimensional displacements are dominated by a small number of components (or dimensions) with very large displacement magnitudes (Appendix C), while measurement error contributes to all of the components equally. Indeed, we find it useful to discard the smallest displacement components from the calculation to improve the signal to noise ratio of the measurement. Specifically, we find that excluding components with displacements < 4$\sigma$ from the calculations effectively removes the effects from noise on $\Delta R$ and $\Delta s$ without significantly perturbing their true scaling and probability distribution exponents, as verified numerically (Appendix C). 

Figure 2b shows that these two experimental high-dimensional measures display a power-law scaling relationship, $\Delta R^2\sim\Delta s ^c$, with $c \approx 1.41 \pm 0.03$. Data corrected for measurement error is shown in green, uncorrected in grey. Such scaling confirms that the configuration path is a fractal with a corresponding fractal dimension $D_f = 2⁄c = 1.42 \pm 0.03$, such that path fragment mass $\sim$ (size)$^{D_f}$. Calculation of the fractal dimension using other methods, such as the correlation dimension \cite{fractalGrassberger1983} shows similar results. Analysis of the configuration space path generated by simulations shows essentially indistinguishable fractal scaling to the experiments (Appendix C).  

The relationship between the droplet super-diffusion and the fractal scaling is straightforward to understand. The configuration path is parameterized by both time $t$ and contour distance $s$. While intermittent dynamics causes $s$ to increase by varying amounts in a given $t$ interval, the corresponding average differences in these variables, $\Delta s$ and $\tau$, are nevertheless proportional,\cite{weeks1998anomalous} Fig. 2c.  This linear correlation indicates that the high-dimensional mean-squared displacement $ \Delta R^2$ will show the same power-law scaling, $\sim \Delta s^c$ and $\sim \tau^c$. Because the individual droplet trajectories are just projections of the configuration space path to lower dimensions, the conventional MSD, $\langle \Delta r^2(\tau) \rangle$, shows the same power-law scaling as well---super-diffusion with the observed exponent satisfying $a \approx c$.

Lastly, the earlier study \cite{hwang2016understanding} also predicted a link between the fractal dimension of the configuration path and a soft glassy material’s power-law viscoelasticity, $G^*(\omega) \sim \omega^\beta$, where $\omega$ is the frequency.  Specifically, if the fluctuating stresses in the material resemble a Brownian random walk, the observed power-law super-diffusion predicts the relation $\beta = D_f^{-1}-0.5$.  This corresponds to $\beta = 0.20 \pm 0.02$ for the experimental $D_f$ and is consistent with direct measurements of the rheology of age-matched emulsions, Fig. 2d, which show $\beta = 0.19 \pm 0.03$. This provides experimental confirmation that the previously unexplained power-law viscoelasticity of SGMs \cite{lavergne2022delayed} is also a result of fractal landscape dynamics.  Conversely, while the fluctuating stresses are not directly measurable in tracking experiments, comparison of the rheology with the tracking data allows us to confirm that the fluctuating stresses resemble the diffusive form seen in simulation.

\section{\label{sec:anisotropy}Anomalous displacements due to landscape anisotropy}

A second anomalous feature of soft glassy dynamics can be seen in the probability distribution of random displacements that occur in a given lag time, Fig. 3a, termed the van Hove self-correlation function. For normal random walks, this distribution has a Gaussian shape. The distribution we find here is distinctly non-Gaussian and heavy tailed---large displacements are much more probable than for a Gaussian distribution with the same width. 

\begin{figure}
  \centering
  \includegraphics[scale=.95]{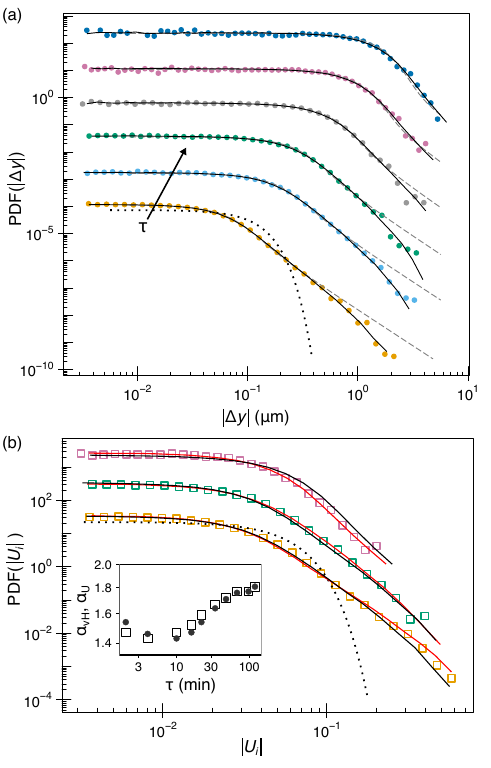}
  \caption{Probability distributions, or van Hove functions, of droplet displacements are non-Gaussian in both real space and configuration space. (a) The van Hove function of individual droplet displacements for $\tau$ = 2, 10, 16, 34, 68, and 120 minutes (bottom to top). Solid black curves represent the best fit ETSD for each $\tau$, dashed curves show the best fit SD, and the dotted curve is a best fit Gaussian distribution for $\tau$ = 2 minutes. (b) The distribution function of the components of high-dimensional displacement unit vectors, $U_i$, for $\tau$ = 2, 16, and 34 minutes (bottom to top). Black curve is a best fit ETSD to the data. The red curves represent a simple construction of $N$ = 775 uncorrelated components with the same ETSD as the data at each $\tau$, and the dotted curve shows a similar construction for $\tau$ = 2 min but with a Gaussian distribution of $N$ = 775 uncorrelated components (isotropic in configuration space). Inset shows $\alpha_{vH}(\tau)$ (circles) and $\alpha_U(\tau)$ (squares).}
  \label{fig:vanHoves}
\end{figure}

Our first task is to determine a suitable fitting function to describe the distribution.  For smaller displacement values, we find that the distributions are well fit by the L\'evy alpha-stable distribution, often called the stable distribution (SD), a family of transcendental functions containing the Gaussian function. The stable distribution is a natural choice for the van Hove of a random process, because it is stable under the repeated convolution corresponding to a generalized random walk\cite{zaburdaev2015levy}. The stable distribution has power-law tails, with an exponent controlled by the stability parameter $\alpha$, and so contains arbitrarily large (positive and negative) values. Distributions of physical variables, however, are typically truncated as very large displacements are physically impossible. Indeed, we find empirically that our van Hove, or displacement distributions are remarkably well fit by an exponentially truncated stable distribution (ETSD), see Fig. 3a. The ETSD satisfies $\textrm{ETSD}(x, \alpha, \lambda) = \mathcal{A} [\textrm{SD}(x, \alpha)\textrm{exp}(-|x|/\lambda)]$, where $\lambda$ is a truncation length and $\mathcal{A}$ is a normalization constant. We find that the observed stability parameter depends on lag time, reaching a minimum of $\alpha_{vH}\approx 1.4$ at intermediate $\tau$ and then trending upward towards $\alpha_{vH} \approx 2$ (Gaussian) at the longest $\tau$, Fig. 3b inset. Such regression to a Gaussian form is expected for any process with a truncated van Hove distribution due to the Central Limit Theorem, but can require a surprisingly long lag time for large $\lambda$ values, \cite{mantegna1994stochastic}.   The small upturn in $\alpha_{vH}$ at short lag time is consistent with the effect of measurement error. Reassuringly, simulation data can also be well fit to the same ETSD form (Appendix D) yielding $\alpha_{vH}$ having similar values and time dependence, see Appendix E.

As previously with super-diffusion, the physical origin of the non-Gaussian van Hove distributions in SGMs is not obvious. A class of literature models predicts such heavy-tailed displacement distributions \cite{cipelletti2003universal, swartz2021active} are due to the $\sim r^{-2}$ dependence of the quadrupolar strain field around a local rearrangement. This gives rise to a truncated power-law tail with $\alpha \simeq 1.5$, which appears inconsistent with our observations. More complicated models with spatially extended, non-quadrupolar deformation fields, however, might lead to a different $\alpha$ value. More generally, L\'evy walk processes are both super-diffusive and can have displacement distributions that resemble stable distributions, but their MSD exponent is related to their displacement distribution \cite{weeks1998anomalous} via $\alpha = 3 - a$, which is also not consistent with the data.  Fractional Brownian motion \cite{mandelbrot1968fractional} is super-diffusive, but has a Gaussian van Hove correlation, $\alpha = 2$ in its simplest realization.  An extension to the fractional Brownian motion model with stable distribution van Hoves has been developed\cite{stoev2004simulation,burnecki2010fractional}, suggesting that similar models might describe our data if truncation were added. 

We have found that the heavy-tailed van Hove correlation is closely related to the anisotropy of the configuration space path, as sketched in Fig. 1b. That is, the directions taken by the configuration space path are not random, as might be supposed. If we consider a set of uniformly distributed points on a unit radius hypersphere (corresponding to random direction unit vectors), their components will be nearly Gaussian distributed in the limit of large dimensionality. This suggests a simple test of random directedness is to compute the high dimensional displacements of the configuration path in a given lag time, to convert them to unit length vectors, and then examine the distribution of their vector components, $U_i(\tau)= (R_i(t+\tau)-R_i(t))/\Delta R(t,\tau)$, pooling values at all $t$. We consider only the $y$ coordinates to calculate $U_i$ since they display the least time-dependent drift. Fig. 3b shows the resulting distribution for our data at three different lag times. The resulting distribution is highly non-Gaussian, demonstrating that the configuration space path is not randomly directed in space; equivalently, this means the valleys in the landscape that the configuration is following are also not randomly directed in space. 

We find that the component distribution $P(U_i(\tau))$ can also be fit by an ETSD form. Indeed, the shape of the distribution function is very similar to that of the van Hove distribution, quantified by the similarity of their stability parameters: $\alpha_U \approx \alpha_{vH}$, Fig. 3b inset.  The nearly constant value of $\alpha_U(\tau)$ for small $\tau$ (which is confirmed in simulation, Appendix E) indicates that this non-random directionality of the configuration path is roughly self-similar on corresponding length scales in configuration space. The observed ETSD van Hove distribution is merely a projection in real space of the distribution controlling the self-similar non-random directionality of the energy landscape valleys. This shows again that a dynamical feature of SGMs in real space is a direct result of a fractal geometrical feature of the energy landscape.  

Na\"ively, we might suppose that the observed non-random directionality in configuration space could be a consequence of correlations between the displacements of different degrees of freedom. However, while the motion of different bubbles must have some finite correlation (due to the affine elastic strain field that connects them), the motion of well-separated bubbles appears nearly uncorrelated.
Indeed, a simple construction shows how non-random directionality can arise {\it without} correlated motion.
Specifically, we can numerically generate an ensemble of $N$-dimensional unit vectors with \textit{uncorrelated} random components having the same ETSD distribution as the observed van Hove distribution. Figure 3b compares the result of of this uncorrelated degree of freedom construction to the experimental $P(U_i(\tau))$, showing very good agreement. This agreement shows that the non-random directionality is not a consequence of correlated motion between droplets, but rather due to the heavy-tailed statistics of individual droplet displacements. Stated another way, when the displacements of individual degrees of freedom are uncorrelated, the $P(U_i)$ and the van Hove distributions have the same shape. 

\section{\label{sec:minima}Avalanches due to fractal clustering of minima}

In the previous sections, we have discussed aspects of the droplet motion in SGMs other than their striking intermittency and cooperativity, to which we now turn. A simple way to quantify cooperativity, when large numbers of droplets move at the same time, is to count how many droplets move more than a threshold amount in a given time interval. Using the displacement truncation length $\lambda$ as a threshold would isolate those droplets undergoing the very largest motions. We choose a lower threshold $\lambda/2 = 0.33 \mu$m for improved statistics, which is still about 10 times larger than our experimental measurement error. Figure 4a shows the number of droplets that move more than that threshold, $N_{\lambda/2}(t)$, in the time interval between consecutive image scans, as a function of time. This function shows large peaks at times when many droplets make large motions. Moreover, a plot of the probability distribution $P(N_{\lambda/2})$ in Fig. 4a(inset) shows a heavy-tailed form, varying as $P(N_{\lambda/2}) \sim (N_{\lambda/2})^{-1.4}$. If every particle moved independently of the others, this distribution would be a peaked binomial distribution. Based on the idea that cooperativity consists of some local droplet rearrangements triggering others, such large, power-law distributed rearrangement events are commonly called avalanches, analogous to those in snow or sand. 

\begin{figure}[ht]
  \centering
  \includegraphics[scale=1]{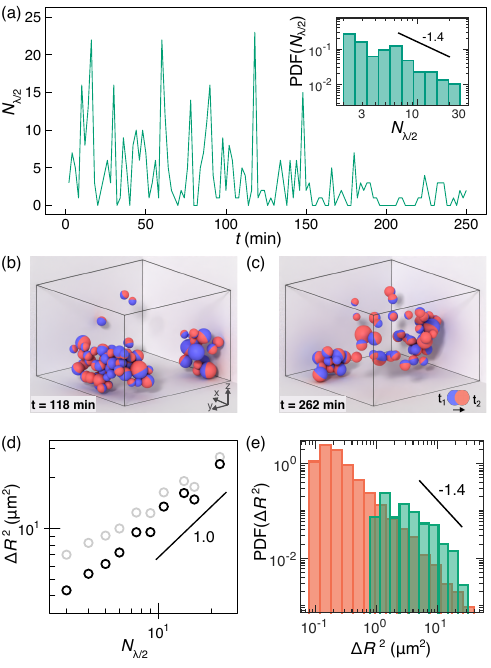}
  \caption{ Intermittent system dynamics at $\tau = 2$ minutes. (a) Number of droplets moving by $\Delta x, \Delta y > \lambda/2$ at each time point. Only data from $t$ = 1 to 150 minutes was analyzed throughout this study. Inset shows the probability distribution of those $t < 150$~min values, following a power law with slope of $-1.4 \pm 0.1$. (b,c) Renders of droplets that move in the top 5\% of all displacements reveal localized clusters during large avalanche events. (d) $\Delta R^2$ for $\tau$ = 2min scales linearly with $N_{\lambda/2}$ (black), data uncorrected for measurement error is in grey. (e) Probability distribution of $\Delta R^2$ also follows a power law with slope of $-1.4 \pm 0.1$ (green), matching that of $N_{\lambda/2}$ and the simulation $\Delta R^2$ (red).}
  \label{fig:avalanches}
\end{figure}

A common method to visualize the spatial arrangement of such avalanches is to prepare a movie that renders only the most mobile droplets, for example, using a threshold such that 5\% of all droplets appear on a time-averaged basis, see Movie A2.  Corresponding images for two typical large avalanches are shown in Fig. 4b-c. To indicate the direction of the droplet motions, the final location of each droplet is rendered in red, the starting location in blue. Because the displacements are small, most droplets render as slightly displaced red and blue hemispheres.
Rendering of simulation data yields similar results (Appendix F). Closer examination reveals the avalanches have a complex spatial structure, forming extended, nearly dense clusters of neighboring droplets. Such clustering is qualitatively similar to the dynamical heterogeneity seen in the cooperative Brownian motion of dense colloidal fluids \cite{weeks2000three}. Further analysis reveals that the number of droplets participating in each cluster follows a power-law distribution, and that the clusters themselves are fractal \cite{weeks2000three} with dimension $D_f \simeq 2.50$ (Appendix F). 

Again returning to the high-dimensional analysis, we seek to understand what features of the energy landscape give rise to these intermittent and cooperative dynamics. Notably, we find that avalanches correspond to large high-dimensional Euclidean displacements, $\Delta R$, between two consecutive points in configuration space. In fact, the two measures of avalanche size are proportional: $\Delta R^2 \propto N_{\lambda/2}$, as shown in Fig. 4d. Since the $N_{\lambda/2}$ values are power-law distributed, this correlation implies that $\Delta R^2$ should be as well; indeed, we find $P(\Delta R^2) \sim (\Delta R^2)^{-1.4}$, shown in Fig. 4e. As expected, small values of $\Delta R$ are dominated by noise in particular when $N_{\lambda/2}$ is small. Applying our error correction method results in improved scaling between the two values and better agreement with the simulation.

Recall that the emulsion relaxes rapidly from one stable energy minimum (where the forces between droplets are in balance) to another, spending most of its time arrested at a minimum. As a result, experimental observations should typically correspond to energy minima. Thus, the measured $P(\Delta R^2)$ between consecutive images may effectively be reporting the distribution of distances between pairs of minima of the energy landscape itself, as sketched in Fig. 1c.  Of course, multiple avalanches may occur between (or during) microscope scans of the sample, so the $\Delta R^2$ between measurements may ‘skip’ some closely spaced minima and undercount small $\Delta R^2$ events.  Examining simulation data confirms this idea.  But importantly, we also find that the limited temporal sampling rate of the experiments does not significantly alter the power-law exponent of the measured $P(\Delta R^2)$, at least for the largest $\Delta R^2$, see Fig. 4e.  

The observed distribution of squared Euclidean distances between minima is unusual, and does not correspond to what would be expected if the minima were randomly distributed along the configuration space path.  Indeed, it is a hallmark of the minima being arranged into fractal clusters along the path, with a fractal dimension that can be derived from the exponent of $P(\Delta R^2)$.  First, given the observed distribution $P(\Delta R^2) \sim (\Delta R^2)^{-1.40}$ and the previously shown relationship between $\Delta R$ and $\Delta s$, it can be shown that $P(\Delta R) \sim \Delta R^{-1.80}$ and $P(\Delta s) \sim \Delta s^{-1.56}$.  To interpret the last scaling form, we can suppose that if the configuration path were stretched out straight, the minima would cluster into a `dust fractal', a fractal with a dimension less than 1. A numerical calculation using the observed exponent for $P(\Delta s)$ indicates that $D_f^{min} \approx 0.5$. See Appendix A for calculation details. In an earlier study, we found that minima are preferentially located in regions of configuration space where the energy landscape is almost flat on longer length scales \cite{hwang2016understanding}, suggesting that here, the fractal clustering of minima is likely yet another manifestation of the underlying fractal structure of the energy landscape itself \cite{massen2007power, charbonneau2014fractal, yoshino2014shear, rainone2015following, biroli2016breakdown, charbonneau2017glass}.

\section{\label{sec:Discussion}Discussion and Outlook}

At the most basic level, the excellent agreement between our experimental measurements and a matched quasi-static simulation confirms the latter model's usefulness for describing dense emulsions and correspondingly, wet foams. At the same time, our experiment confirms that fractal landscape dynamics is the origin of the previously mysterious soft glassy dynamics in those materials. In particular, we have found nearly one-to-one correspondence between real-space observables, including the mean squared displacement, $\left<\Delta r^2_{xy}(\tau)\right>$, the van Hove correlation function $P(\Delta y(\tau))$ and the avalanche number $N_{\lambda/2}(t)$ with three measures of the high-dimensional self-similar geometry, $\left<\Delta R^2(\tau)\right>$, $P(U_i)$ and $\Delta R^2(t)$ of the configuration path and its minima. More generally, this study demonstrates that the physically relevant geometrical features of the high-dimensional energy landscape are not visible only in a simulation or an analytical calculation, but can be reliably deduced from multi-particle tracking data. While we have thoroughly characterized the fractal geometry of the energy landscape that emerges in a dense emulsion near jamming, the mathematical origin of this geometry remains mysterious.  The appearance of similar dynamics in other systems is suggestive that such landscapes may emerge in a variety of other systems.  Future work will seek to understand the emergence of such fractal geometry, the effect of material properties like viscosity on fractal landscape dynamics, and to develop a dynamic model for the random motion and intermittency in these materials.

We expect the exploration of high-dimensional landscapes from empirical data may prove useful in a variety of systems with similar dynamics, such as cytoskeletal networks \cite{hoffman2009cell} and perhaps even neural networks \cite{baity2018comparing, chen2022anomalous}. High resolution multi-particle tracking data in cells \cite{shi2019dissecting, shi2021pervasive} may enable the characterization of the cytoskeleton’s energy landscape, enabling the screening or refinement of emerging cytoskeletal models. Practical applications of AI rely on deep learning, where computationally costly learning processes are accelerated by ‘shortcut’ connections \cite{lecun2015deep} in the network, which alter the structure of the high-dimensional ‘loss’ landscape. Our analysis may lead to a clearer understanding of deep learning dynamics \cite{chen2022anomalous}, and more efficient learning algorithms.

\begin{acknowledgments}
We are grateful for useful conversations with Doug Durian, Fran\c{c}ois Lavergne, Andrea Liu, Christopher Porter, Yu Shi, Talid Sinno,  Veronique Trappe and Eric Weeks. We are also grateful to Dr. Bomyi Lim for confocal microscopy and Dr. Paulo Arratia for rheometry.  This work was supported by NSF-DMR 0706388, 1609525, and 1720530, NSF-PHY 1915193 and 1915174 and the David and Lucile Packard Fellowship, with computational resources provided by XSEDE through TG-DMR150034.

CRC, MM, AT, RAR and JCC designed research, analyzed the data and wrote the paper. CRC collected the data. AT performed the simulations. KF and VNM formulated the experiment and collected preliminary data. SS and DHR contributed to data analysis and interpretation.
\end{acknowledgments}


\appendix

\section{Materials and Methods}

\subsection*{Sample preparation}
The O/W emulsion was prepared by slow, dropwise addition of the dispersed phase (80\% v/v) to the continuous phase with constant homogenization (IKA T18) at 21,500 rpm. The continuous phase contained 3\% (w/w) Synperonic PE P105 (Sigma-Aldrich) surfactant dissolved in a mixture of 95\% (w/w) formamide, 5\% (w/w) water. For confocal imaging fluorescein sodium salt was dissolved in the water component at 2.7 mM concentration, prior to mixing and emulsification. The dispersed phase contained a mixture of 94\% (w/w) 1-bromohexane and 6\% (w/w) octane. Following emulsification, the sample was centrifuged for 10 minutes at 700 rpm for removal of air bubbles, and was aged in a closed microscopy chamber at room temperature. This chamber consisted of stacked \#1.5 coverslips used as spacers, topped by a \#0 coverslip and sealed with high viscosity UV glue (Norland 68T).  The dense emulsion was then imaged using a Zeiss LSM 800 confocal microscope with an oil immersion objective.  The rate of ripening in the emulsion is seen to decrease slowly over time, until nearly complete arrest occurs 10-11 hours after formation.  We conjecture this is due to the increase of surfactant concentration in the connected phase, and a corresponding decrease in droplet surface tension.  All analyses are performed for $t < 150$~mins or 7-9.5 hours after emulsion formation.

\subsection*{Rheology} Measurements were performed using a strain-controlled rheometer (DHR-3 TA Instruments) with a parallel plate geometry (40 mm plate diameter). All measurements were done in the linear viscoelastic regime (strain $\epsilon = 1\%$), which was verified by an amplitude sweep at $\omega = 1$ rad/sec. Complex shear modulus values were obtained from frequency sweeps at $T= 25 \pm 1^\circ$C and frequencies from $\omega = 0.016$ to $30$ rad/sec. The gap size ranged from 90-150 $\mu$m and the same sample was measured multiple times to observe the effects of emulsion aging. Frequency scans in both directions (from low to high $\omega$ and vice versa) showed similar results and were averaged together. Measurements were started 7 hours after emulsion preparation to reach dynamical scaling and to correspond to the confocal imaging data.

\subsection*{Emulsion Simulation} We simulate our dense emulsion using a modified 3-D bubble model, extending the one used in our previous study \cite{hwang2016understanding}, based on a system of polydisperse soft-spheres at a volume fraction $\phi = 0.75$, with pairwise interaction energy:
\begin{equation}
 V(\mathbf{d}_{ij})=\begin{cases}
    \frac{\epsilon}{2} {\left(1 - \frac{\lVert{\mathbf{d}_{ij}}\rVert}{r_i + r_j} \right)}^{2}, & \text{if $\lVert \mathbf{d}_{ij} \rVert<r_i + r_j$}\\
    0, & \text{otherwise},
 \end{cases}
\label{eq:soft-sphere potential}
\end{equation}
$\mathbf{d}_{ij}$ being the distance between two bubbles (soft-spheres) of radii $r_i$ and $r_j$. The bubbles exchange mass due to differences in notional Laplace pressure according to:
\begin{multline}
    Q_i=-\alpha_1 \sum_{j}^{neighbors} (\frac{1}{r_i}-\frac{1}{r_j})A_{overlap}-\alpha_2(\frac{1}{r_i}-\frac{1}{<r>})r_i
\end{multline}
 
The evolution of the system is considered in the quasi-static limit - where the energetic relaxation time is much smaller than the ripening time scale. This leads us to relax the system to a minimum between consecutive ripening moves. The parameters for the simulation are similar to Ref. \cite{hwang2016understanding}.
The system is initialized using a Gaussian distribution of bubble radii, and its properties are considered once the system reaches a dynamical scaling state. Under such a steady state, the droplets reach a radii distribution resembling a Weibull distribution, $P(r) = (k/ \lambda) (r/ \lambda)^{k-1}\textrm{exp}(-(r/\lambda)^k)$, where $k \approx 1.66$ and $\lambda$ is a scale parameter.

\subsection*{Fractal Exponent Calculation} In general, if a random variable $x$ is power-law distributed, $P(x) \sim x^a$, and a second variable $y$ scales as $y \sim x^b$ then $y$ is also power-law distributed with $P(y) \sim y^c$ and $c=(a+1)/b -1$.  This relation is used to estimate the scaling exponents for $P(\Delta R)$ and $P(\Delta s)$ from those for $P(\Delta R^2)$ and the scaling exponent $c \approx 1.41 \pm 0.03$ defined in the main text.  To estimate the dust fractal dimension of the minima in $s$, we created simple asymmetric L\'evy walks \cite{weeks1998anomalous} by cumulative summing uncorrelated positive random numbers $x$ having a power-law distribution of values, $P(x) \sim x^{-d}$, and then computing their fractal dimension $D_f^{min}$ using a correlation dimension \cite{fractalGrassberger1983}.  The results could be well fit by the empirical form: $D_f^{min} \approx ((1/(d-1)^3)+1)^{-1/3}$, which was used to compute the experimental value.

\section{Dynamical Scaling State}
\begin{figure}[ht]
\centering
\includegraphics[scale=.95]{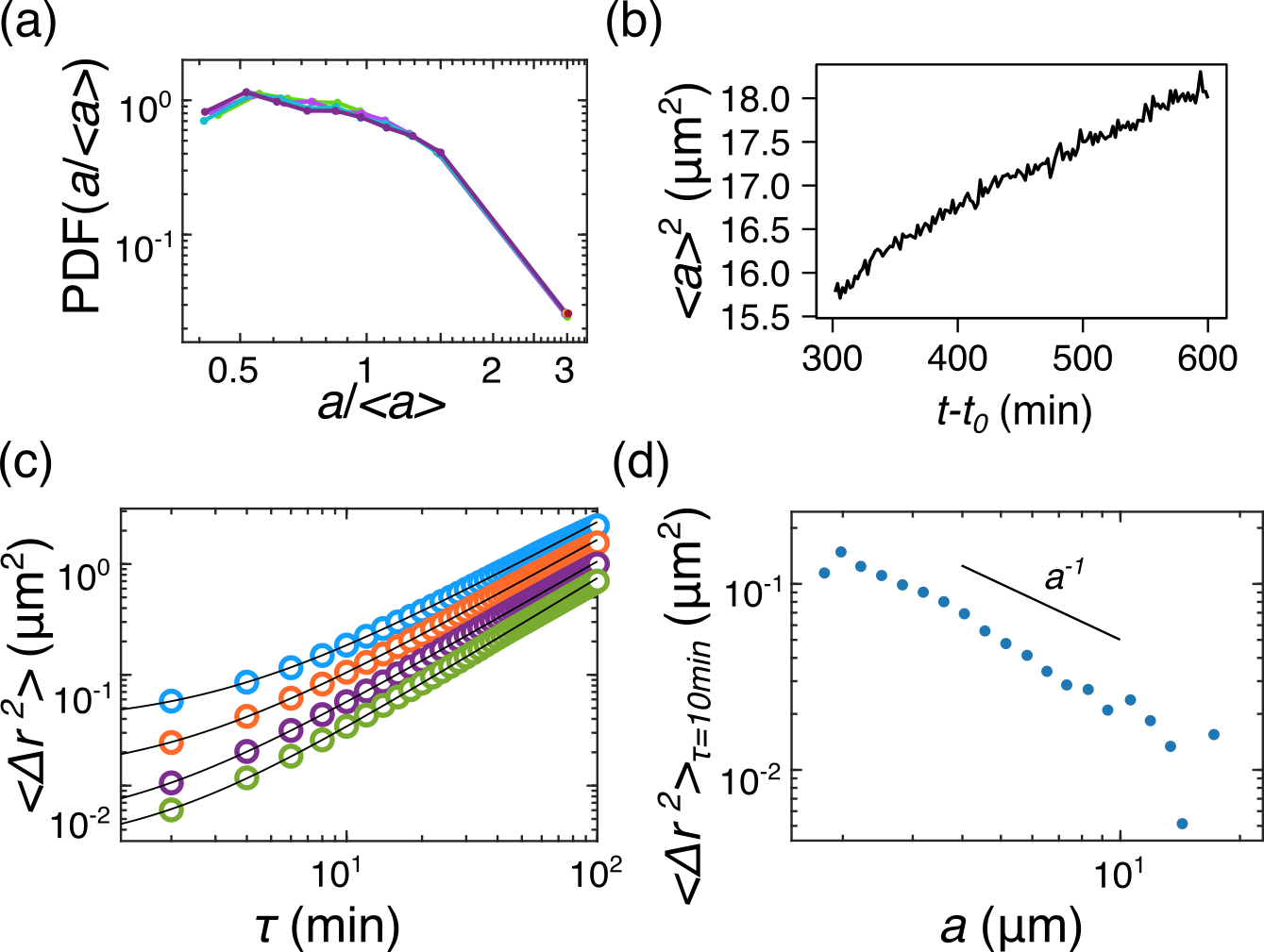}
\caption{Experimental system agrees with expected dynamical scaling behavior and polydispersity effects. (a) Droplet size distribution evolution throughout the dynamical scaling period, normalized by the average droplet size, where $t-t_0$ represents time from the beginning of data collection and $t_0 = 7$ hrs is the time from emulsification. (b) Time evolution of the squared average droplet radius shows expected linear relationship. (c) Mean squared displacement of droplet ensembles grouped by radii: [1.60-2.65] $\mu$m (blue), (2.65-3.79] $\mu$m (orange), (3.79-5.33] $\mu$m (purple), (5.33-18.21] $\mu$m (green). Black curves show fits to a power law plus a constant $y=ax^b+c$, where $c=2\sigma^2$ and $\sigma$  represents measurement error. The displacements of smaller droplets contain larger error, as expected, due to image resolution limitations. (d) Mean squared displacement at $\tau = 10$ min as a function of droplet radius follows the expected $a^{-1}$ behavior.}
\end{figure}

Figure S1 confirms that the system has reached a dynamical scaling state, where the squared droplet size increases linearly with time and the shape of the size distribution does not change significantly (Figs. S1a,b). As expected for droplets in a mechanical continuum, larger droplets diffuse more slowly (Figs. S1c,d), showing the same scaling as the Stokes-Einstein relation despite being driven by active stress fluctuations.

\section{Measurement Error Correction}

\begin{figure}[ht]
  \centering
  \includegraphics[scale=.90]{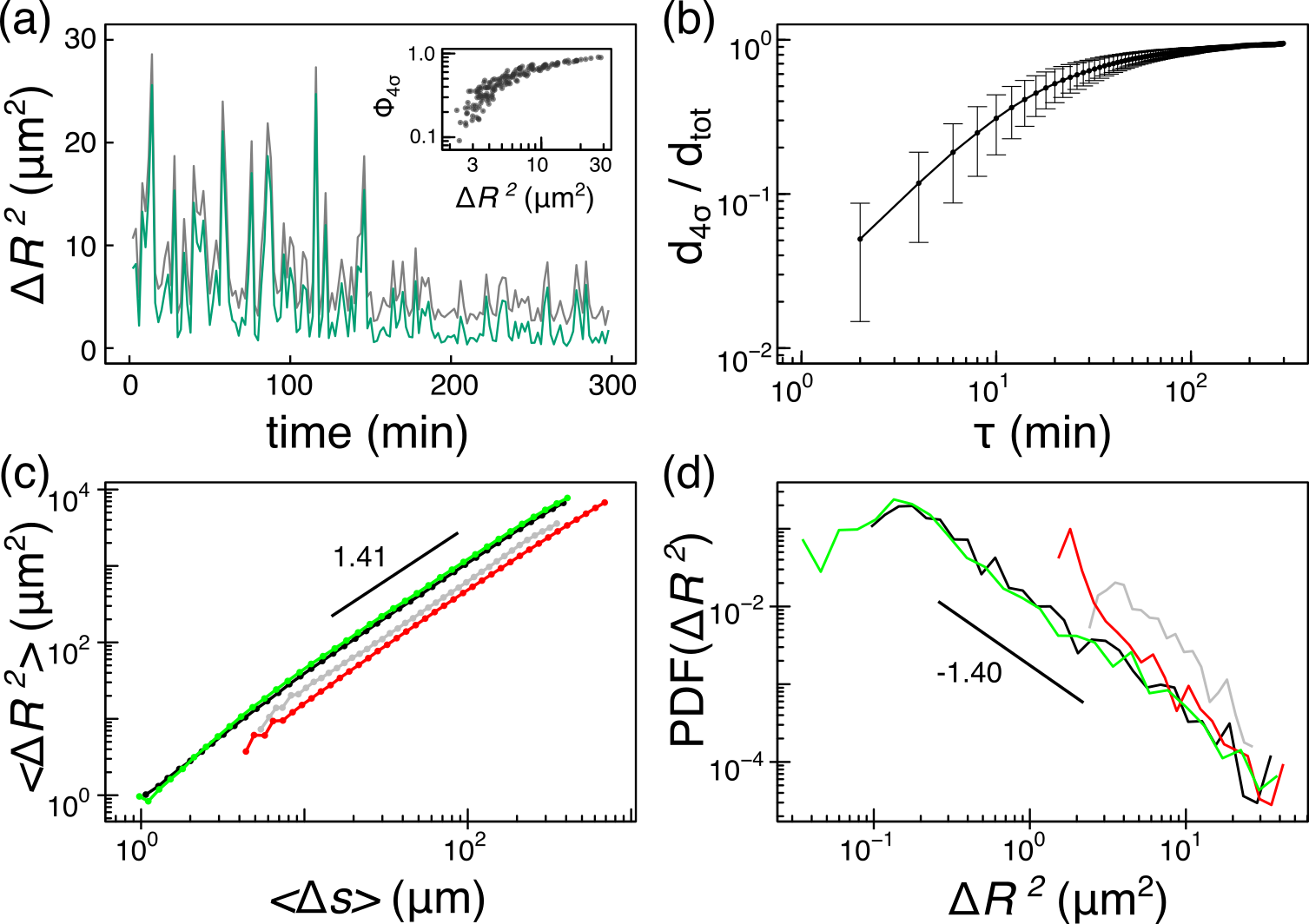}
  \caption{Gaussian noise effects on high-dimensional displacements. (a) Experimental $\Delta R^2$ before (grey) and after (green) removing measurement error. Inset shows the fraction of droplets moving $> 4\sigma$ for each $\Delta R^2$ value. (b) Fraction of droplets moving $> 4\sigma$ for all $\tau$ values. (c) $\Delta R^2$ and $\Delta s$ between pairs of simulation configurations (black), simulation after adding Gaussian noise (red), and simulation with Gaussian noise after using the noise removal method described in the text (green). Experiment results before removing measurement error are shown in grey. (d) Probability distribution of $\Delta R^2$ for $\tau = 1$ simulation time step and $\tau = 2$ min in the experiment. Color scheme is the same as (c).}
\end{figure}

The experimental measurement error can be quantified by fitting the mean-squared displacement of individual droplets to a power law plus a constant, $MSD_{xy}=A\tau^B+C$, where $C = 4\sigma^2$. The fit is shown in Fig. 2a (main text), where $4\sigma^2 = 0.0036$ $\mu$m$^2$ and $\sigma = 0.03$ $\mu$m. This presumably perturbs high-dimensional displacement calculations, especially for low values of $\tau$. To show how random error affects our data, we added a Gaussian-distributed noise signal with zero mean and $\sigma = 0.03$ $\mu$m to the noise-free simulation data (re-scaled for comparison to the experiment). As shown in Figs. S2c-d (red data), Gaussian error significantly alters the original simulation results by omitting the smallest displacements.

To reduce the sensitivity of our experimental analysis to measurement error, we modified the calculation of all high-dimensional Euclidean distances to exclude any contribution from components/dimensions that were below a threshold. We found empirically that a threshold of $4\sigma$ was optimal for the analysis of noisy data to nearly revert to that of the original noiseless data, see Figs. S2c-d (green data). For the distribution of $\Delta R^2$ at $\tau = 1$ simulation time step, noise completely changes the shape and slope of the distribution. This modified calculation was therefore applied to the experimental $\Delta R^2$ and $\Delta s$ values throughout our analyses in order to reduce systematic errors due to noise in the results. Fig. S2a shows the result of this error correction on the experimental $\Delta R^2$, bringing the noise floor closer to zero, while the inset shows that only a very small fraction of droplets moves $> 4\sigma$ at the smallest $\Delta R^2$ values. This confirms that the smallest $\Delta R^2$ values are most affected by noise, compared to larger values which are dominated by large droplet displacements. Moreover, data at shorter $\tau$ values is also more severely affected by noise as shown in Fig. S2b, where the fraction of droplets moving $> 4\sigma$ approaches 1 at large $\tau$.

\section{Emulsion Simulation Displacements}

\begin{figure}[ht]
\centering
\includegraphics[scale=1]{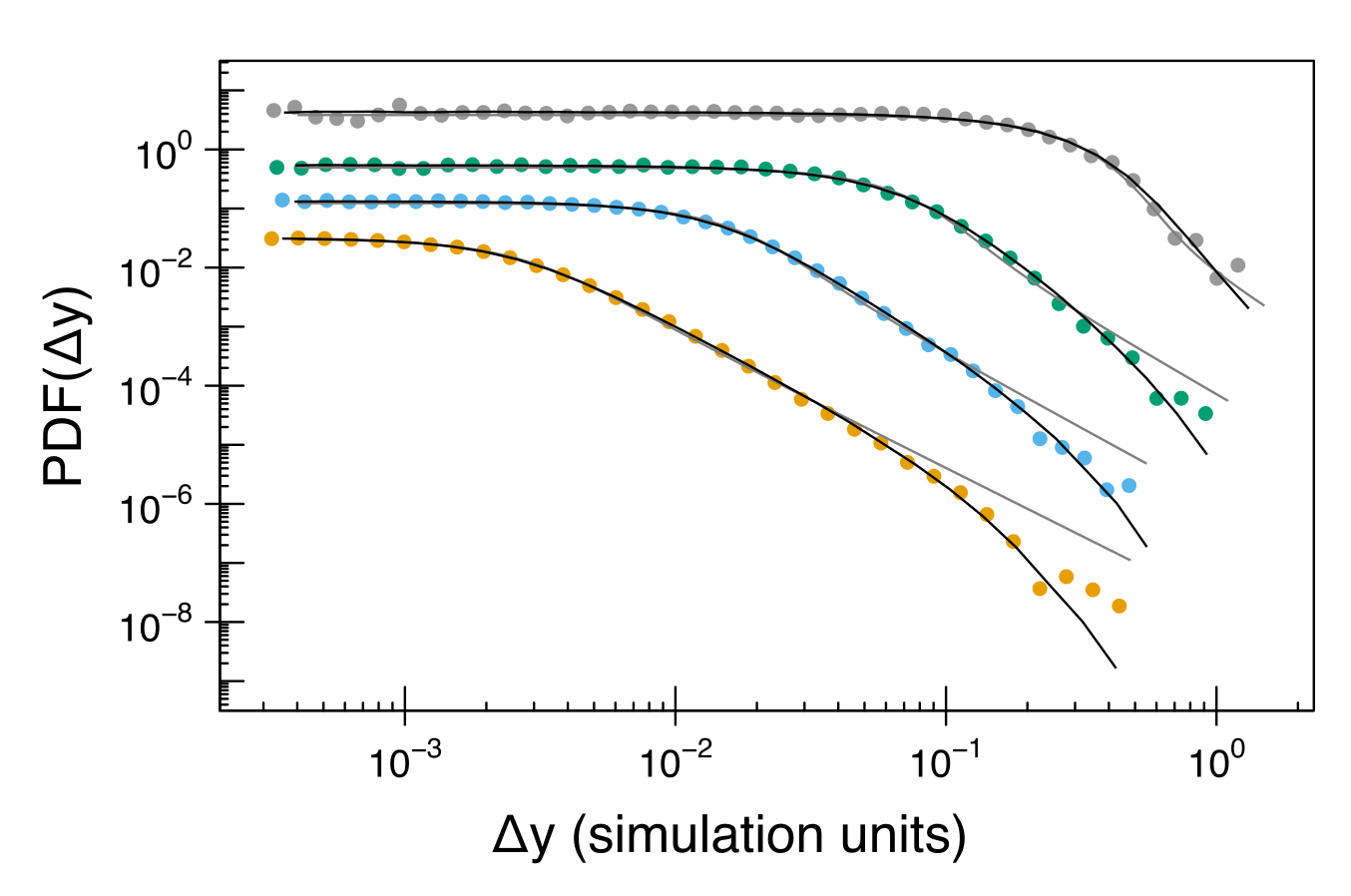}
\caption{Van Hove function of simulation droplet displacements for $\tau$ = $1$, $5$, $24$, and $140$ simulation time units (bottom to top). Solid black curves represent the best fit ETSD and grey curves show the best fit SD.}
\end{figure}

Figure S3 shows the distribution of droplet displacements in simulations, for multiple values of $\tau$. These are well fit by an ETSD, and the $\alpha$ values from the fits follow a time-varying trend similar to the experimental data (see Fig. S4).

\section{ETSD Shape Parameter}

\begin{figure}[ht]
  \centering
  \includegraphics[scale=1.2]{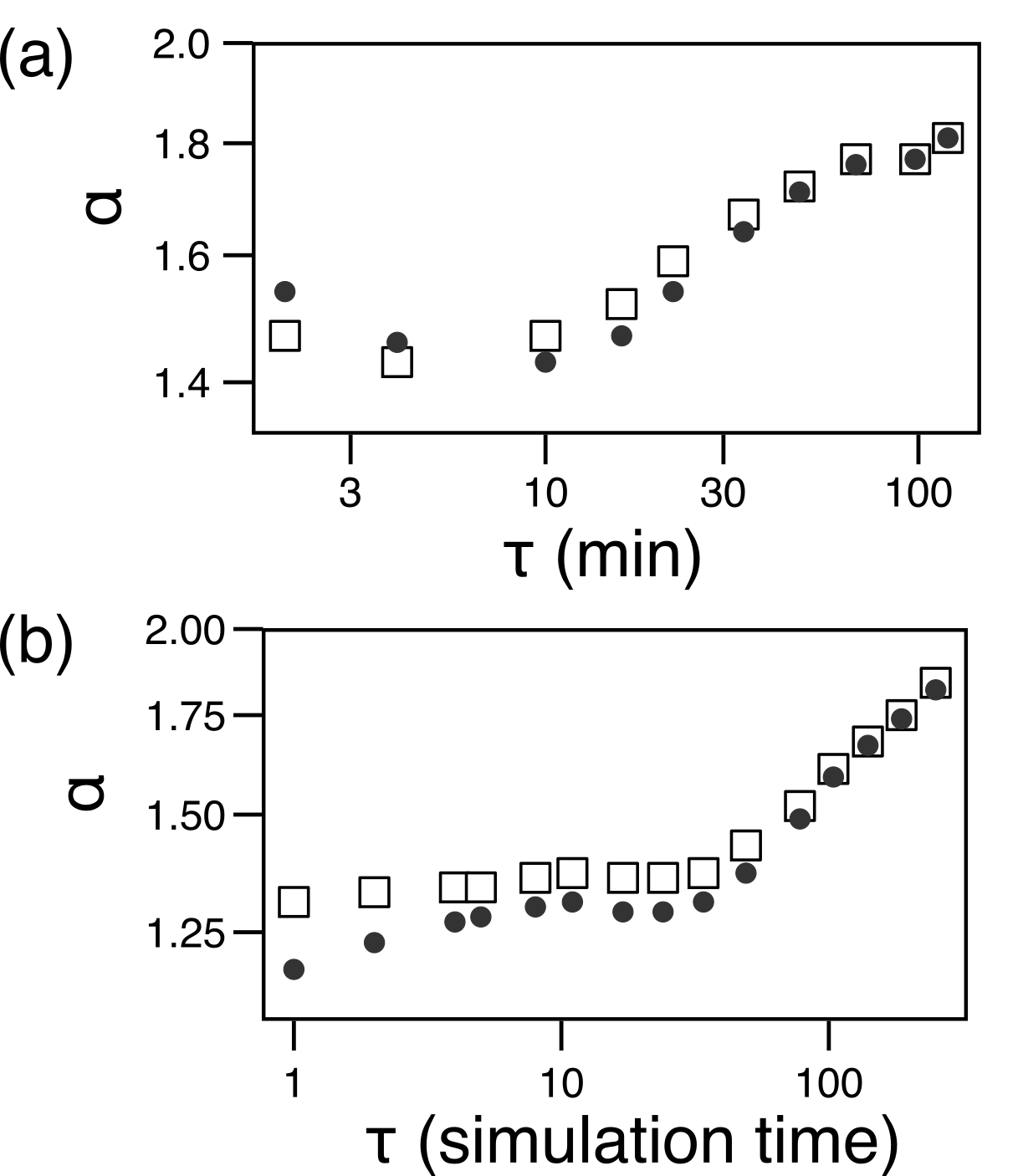}
  \caption{ETSD stability parameter for the lag-time dependent van Hove displacement distribution, $\alpha_{vH}$ (circles), and for the high-dimensional displacement vector components distribution, $\alpha_U$ (squares), shown for the emulsion experiment (a), and emulsion simulation (b).}
  \label{fig:S9}
\end{figure}

We have found that an exponentially truncated stable distribution (ETSD) provides a useful fitting form for the van Hove distribution in systems with fractal landscape dynamics, as shown in Figs. 3 and S3. The stability parameter $\alpha$ from those fits provides a measure of how heavy-tailed the distribution is, related to the exponent of the power-law tail in the untruncated SD.  These $\alpha$ values, $\alpha_{vH}$, show a non-trivial $\tau$ dependence shown in Fig. S4, closely resembling that of the high-dimensional displacement vector components, $\alpha_U$.  Both experiment and simulation appear to be trending to a Gaussian value $\alpha=2$ at long times, due to regression according to the Central Limit Theorem. The near constant value of $\alpha_U$ in the simulation case for small and intermediate lag times confirms the self-similarity of the non-random path directions in configuration space.

\section{Avalanche Clusters}

\begin{figure}[hb]
  \centering
  \includegraphics[scale=1.1]{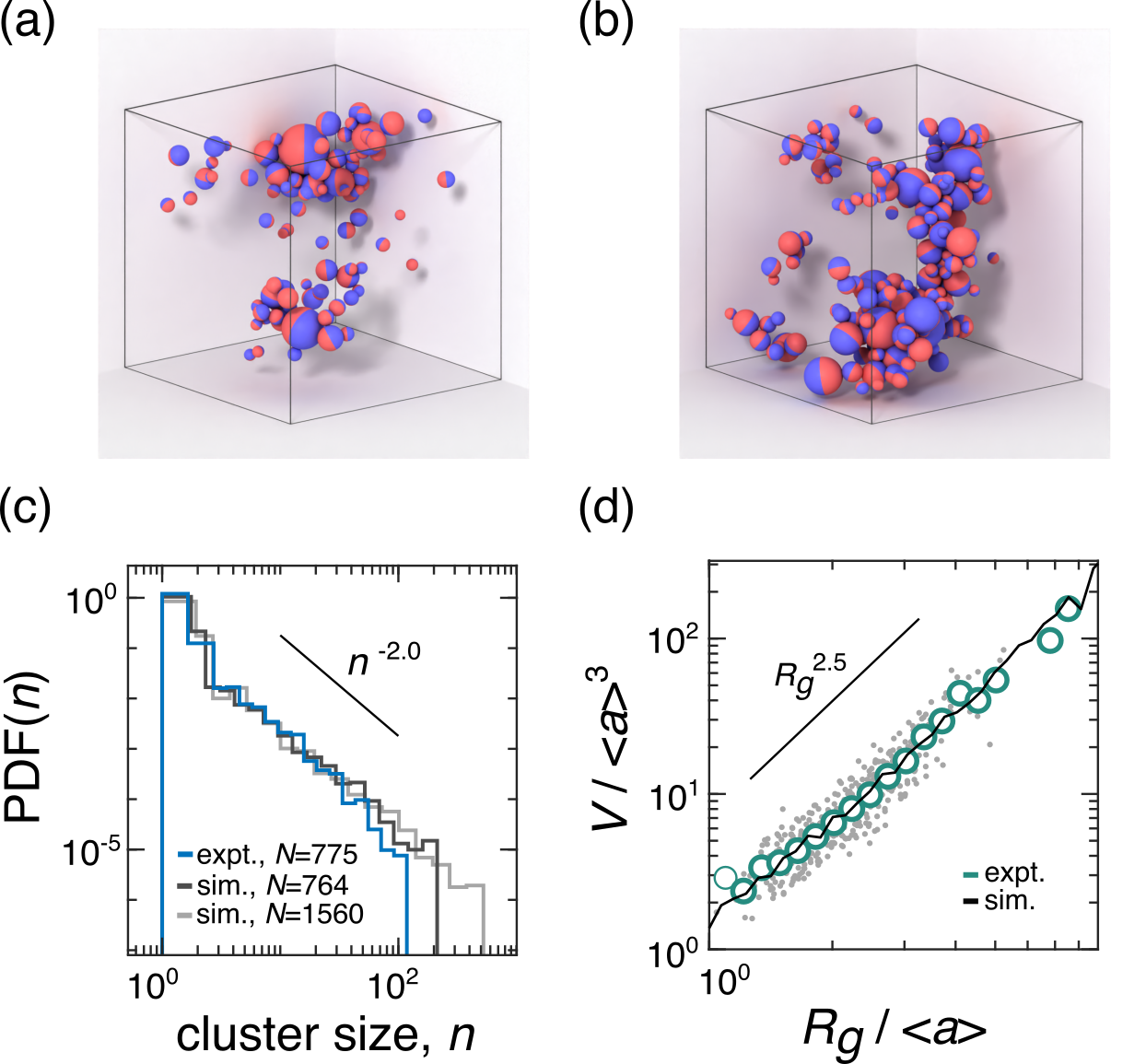}
  \caption{Cluster analysis for droplets with the largest displacements. (a,b) Simulation data renders of the particles moving in the top $5\%$ of all particles at $2$ different time points, calculated with the same experimental threshold method described in the text. (c) Distribution of the number of droplets, $n$, in each cluster shows power law scaling for both experiment and simulation. (d) The volume of individual clusters, $V$, shows a power law dependence on their radius of gyration, $R_g$, confirming their fractal shape and matching the averaged simulation results (line). Green circles are averages of the grey data points.}
\end{figure}

The droplets with the largest displacements between consecutive confocal images ($\tau = 2$ min) were determined by using a time-dependent threshold, constructed so that 5$\%$ of droplets were above threshold on a time averaged basis. These particles were formed into clusters using an adjacency matrix that specifies which droplets are contacting neighbours. Droplets were considered to be in contact if their center-to-center separation was less than $1.1$ times the sum of their radii, to allow for measurement error and droplet distortion.  Our findings were not sensitive to this factor. The resulting clusters from the experiment show fractal scaling (Figs. S5c-d) and are similar to those observed in the simulations (Figs. S5a-b).

\bibliography{arXiv/arXiv}

\end{document}